\documentclass[%
 reprint,
 amsmath,amssymb,
prd,
floatfix,
]{revtex4-1}
\usepackage{float}
\usepackage[utf8]{inputenc}
\usepackage{graphicx}
\graphicspath{ {./figs/} }
\usepackage{dcolumn}
\usepackage{bm}
\usepackage[mathlines]{lineno}
\usepackage{xcolor}
\usepackage{soul}
\usepackage{bm}
\usepackage{bbm}
\usepackage{amsmath,amsfonts,amssymb}
\usepackage{extarrows} 
\usepackage{mathtools}
\usepackage{hyperref}
\hypersetup{
colorlinks,citecolor=red,
filecolor=blue,
linkcolor=blue,
urlcolor=black
}

\newcommand{\nc}{\newcommand}
\nc{\nn}{\nonumber}
\nc{\txt}{\textrm}
\nc{\txtsup}{\textsuperscript}
\nc{\txtsub}{\textsubscript}
\nc{\calL}{\mathcal{L}}
\nc{\U}{\mathcal{U}}
\nc{\T}{\mathcal{T}}
\nc{\E}{\mathcal{E}}

\begin{document}


\title{Signatures of discrete time-crystallinity in transport through an open Fermionic chain}

\author{Subhajit Sarkar }
\email{subhajit@post.bgu.ac.il}
\affiliation{Department of Chemistry, Ben-Gurion University of the Negev, Beer Sheva, 84105,  Israel}
 \affiliation{School of Electrical and Computer Engineering, Ben-Gurion University of the Negev, Beer Sheva, 84105,  Israel}

\author{Yonatan Dubi}
\email{jdubi@bgu.ac.il}
\affiliation{Department of Chemistry, Ben-Gurion University of the Negev, Beer Sheva, 84105,  Israel
}
\affiliation{Ilse Katz Center for Nanoscale Science and Technology, Ben-Gurion University of the Negev, Beer Sheva, 84105,  Israel
}

\date{\today}

\begin{abstract}
\begin{center}
    \textbf{Abstract}
\end{center}
Discrete time-crystals are periodically driven quantum many-body systems with broken discrete-time translational symmetry, a non-equilibrium steady state representing self-organization of motion of quantum particles. Observations of discrete time-crystalline order are currently limited to magneto-optical experiments. Crucially, it was never observed in a transport experiment performed on systems connected to external electrodes. Here we demonstrate that both discrete time-crystal and quasi-crystal survive a very general class of environment corresponding to single-particle gain and loss through system-electrode coupling over experimentally relevant timescales. Using dynamical symmetries, we analytically identify the conditions for observing time-crystalline behavior in a periodically driven open Fermi-Hubbard chain attached to electrodes. Remarkably, the spin-polarized transport current directly manifests the existence of a time-crystalline behavior. Our findings are verifiable in present-day experiments with quantum-dot arrays and Fermionic ultra-cold atoms in optical lattices.
\end{abstract}

\maketitle

\section*{Introduction}
Spontaneous symmetry breaking represents a unifying concept which ubiquitously spans from condensed matter and atomic physics to high energy particle physics \cite{Anderson393}. Examples include, superconductors, Bose-Einstein condensates, (anti)ferromagnets, all the crystals, and even (Higgs) mass generation for fundamental particles \cite{BCS_PhysRev1957, Anderson198, BEC_Ketterle, chaikin_lubensky_1995, cottingham_greenwood_2007}. However, time-translation symmetry has always been special: Sch\"{o}dinger's equation, which governs quantum physics, is indeed time-translation invariant. In spite of that,  time-translation symmetry breaking has been shown to be possible under special circumstances, leading to the discrete time-crystal (DTC) behavior. Periodically driven (Floquet) closed quantum systems that never reach a thermodynamic equilibrium can indeed exhibit a DTC behavior \cite{  Sacha_2017, Khemani_PRL_DTC,Else_PRL_Floquet_DTC, Surace_prb_DTC, khemani2019brief, Wilczek_prl_quantum_tc, Choi2017, Zhang2017}. 
 The breaking of discrete time translation symmetry is manifested in the sub-harmonic oscillations of the order parameter and typically rely on disorder and localization to avoid reaching a stationary state of infinite temperature that opposes time crystalline order \cite{Yao_Nayak_TC_physics_today, bukov_polkovnikov_flloquet_2015,Khemani_PRL_DTC, Moessner2017}. The first signature of DTC was reported in strongly interacting spin systems \cite{Choi2017, Zhang2017}, measured using  spin-dependent fluorescence (optical measurements) \cite{Choi2017, Zhang2017,Kyprianidis1192}. Subsequently, DTCs have been realized in variety of systems \cite{Dogra_science.aaw4465, pal_prl_cluster, magnon_space_time_crystals_prl, smits_prl_superfluid_QG,Autti2021}.
 Recent advances in the fabrication and control of exchange coupled quantum-dot  arrays \cite{Takumi_APL_2018, Mukhopadhyay_APL_2018, Mills2019, Sigillito_PhysRevApplied2019, Qiao_PhysRevX_2020} led to the theoretical proposal of observing a DTC \cite{Economu_PhysRevB_2019} and subsequent experimental realization of the same leading to stable quantum information processing \cite{Qiao2021, vandyke2020protecting, EstarellasSci_Adv8892}.
 
 Observations of DTCs has been limited to closed Floquet quantum systems (i.e., periodically driven quantum system without dissipation), as a transient phenomenon restricted to optical detection from spin-dependent fluorescence or magnetization measurements \cite{Choi2017, Zhang2017, Kyprianidis1192, mi2021observation}. Consequently, in spite of the recent observation of DTC in various platforms \cite{Choi2017, Zhang2017,Kyprianidis1192, pal_prl_cluster, magnon_space_time_crystals_prl, smits_prl_superfluid_QG, Qiao2021, vandyke2020protecting} the  {experimental observations of} DTCs in dissipative quantum many body systems rarely has been  {reported} until recent observation in the atom-cavity system \cite{Dissipative_DTC_PhysRevLett_atom_cavity}. 

 Opening the system to an environment indeed presents a challenge of defining the notion of DTC. Nevertheless, plausible criteria, based on the spectrum of the {Floquet map} (i.e., the one period time evolution operator), to define and characterize DTC, has been put forward  \cite{RieraCampeny2020timecrystallinityin, Gong_PhysRevLett} in open systems governed by a Markovian Lindblad master equation \cite{Lindblad, GKS, Breuer, Archak_Dhar_Kulkarni_PhysRevA}. It relies on the existence of several non-decaying states of the {Floquet map} with eigen-values $\mathcal{E}_{DTC} \in \lbrace \mathcal{E}_{\mu} \rbrace$ such that $\mathcal{E}_{DTC} \neq 1$ but $(\mathcal{E}_{DTC})^{p} =1$ for some integer `$p$' \cite{RieraCampeny2020timecrystallinityin}.  {Coupling to the external environment makes the DTCs fragile \cite{Lazarides_Moessner_2017, Choi2017, Zhang2017}, but dissipation engineering has shown the promise. In this regard, the mechanisms that stabilize DTCs in open quantum systems can be classified into three broad categories. These are, mean-field based DTCs which are stable only in the thermodynamic limit \cite{BTC_PhysRevLett_Fazio,carollo2021exact, Gong_PhysRevLett, Marino_Demler_NJP_2019}; symmetry-based mechanism \cite{Buca_nat_comm2019, diss_Flq_DTC_PRL_Ikeda}; and meta-stable DTCs that neither require any symmetry nor disorder \cite{Gambetta_PhysRevLett_metaDTC}. }
 
  The mean-field-based DTCs require a well-defined semi-classical limit and are found to be fragile to quantum fluctuations \cite{Gong_PhysRevLett, Marino_Demler_NJP_2019, Cosme_2020, RieraCampeny2020timecrystallinityin}. The symmetry-based mechanism can accommodate quantum fluctuations. However, it requires a very specific choice of system-environment coupling that satisfies the existence of dark state space to stabilize DTCs in a dissipative quantum-many body system \cite{diss_Flq_DTC_PRL_Ikeda, Buca_nat_comm2019, Buca_prb_2020, Buca_PhysRevLettQG_2019, Prosen_SciPostPhys_2020, Tindall_2020}. Indeed, as detailed in Ref.~\onlinecite{Buca_nat_comm2019}, dissipation (in the form of on-site dephasing) is essential for stabilizing the DTC behavior, because it allows for coupling (and hence synchronization) of different sectors in the Floquet Hamiltonian which would otherwise remain decoupled in absence of dephasing.

  However, most of the realistic environment - specifically, a transport setup in which the system is open to charge transfer from electrodes - does not satisfy the dark state criterion. How DTCs survive or die out due to coupling with this type of environment is currently unknown. We show that a DTC and discrete-time quasi-crystal (DTQC) \cite{Pizzi_PRL_DTC, DTQC_PhysRevB_Sacha, Zhao_DTQC_PhysRevB} can indeed survive a very general class of environment, viz., the single-particle transport through system-electrode coupling, over a sufficiently long time when the system environment coupling is weak. With stronger system-lead coupling, the system reaches a set of transient states distinct from the usual Floquet steady-state (FSS). 
  Strikingly enough, the spin-polarized transport current through the paradigmatic Floquet Fermi-Hubbard chain connected to electrodes can manifest DTC/DTQC order, linking the DTC physics with electrical transport.
 
 In the following, we first establish, supported by numerical calculations, that even if the system is connected to external electrodes (leads), there emerges a unique ``weak local symmetry," to be described later, that remains preserved as long as the system-lead coupling is weak. The DTC (and the DTQC) manifests itself in a sharp peak of the Fourier transform of the expectation value of the spin-current that respects the ``weak local dynamical symmetry". The temporal oscillation of the spin-current is locked at a frequency which provide a direct measure to the DTC (and the DTQC) sub-harmonic frequency. For strong system-lead coupling both DTC and DTQC indeed decay, however, for an intermediate strength the system reaches a (possibly degenerate) transient manifold \cite{Macieszczak_PhysRevLett_Metastability} and show long lived (and arguably pre-thermal \cite{Kyprianidis1192, mi2021observation}) DTC(DTQC) behavior before it decays. To this end we clarify that the notion of transient manifold we use pertains to the finite lifetime of the DTC, in contrast to the meta-stable manifold corresponding to the vanishing gap of the spectrum of Liouvillian/ Floquet map \cite{Gambetta_PhysRevLett_metaDTC}. We further show, from the spectrum of the Floquet map, that the decay of DTC (DTQC) scales linearly with the system-electrode coupling with the slope being twice the driving period. The phase of DTC eigenvalues of the Floquet map further shows the frequency locking phenomena observed in the oscillating spin-current. This frequency locking is a consequence of the emergence of the weak local dynamical symmetry. The scaling behavior and frequency locking phenomena we obtain is independent of the values of the system parameters, the magnitude of the external drive, the choice of the initial state, and system size. 
 \section*{Results and Discussions}
\subsection*{System}
\begin{figure}
 \centering
 \includegraphics[keepaspectratio=true,scale=0.55]{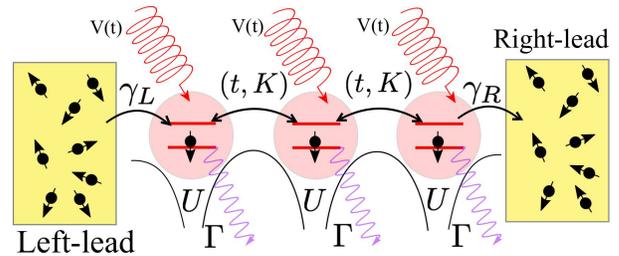}
\caption{{ \bf Schematic arrangement of the quantum-dot array set-up.} $t,U,~\text{and}~K$ are the hopping, onsite and nearest-neighbor interactions, respectively. $V(t)$ represents the external time-periodic drive. $\gamma_{L}$ and $\gamma_{R}$ are the system left-lead and system right-lead couplings, respectively (note the direction of the arrow indicating the one-way electron transfer, i.e., an infinite bias condition) which is taken to be the same for both spin-up and spin-down electrons. Onsite dephasing is given by $\Gamma$.
 }
 \label{fig_set-up}
 \end{figure}

In order to be specific, we consider a quantum-dot array consisting of $N$ dots (numerical simulations are done with $N=3$) attached to external electrodes, see Fig. \ref{fig_set-up}. Such a quantum-dot array set-up can realize a one-dimensional Fermi-Hubbard model \cite{Hensgens2017}. However, our results are equally valid for Fermi-Hubbard model routinely realized in optical lattices \cite{Gross995}. Each site of the Fermi-Hubbard chain is irradiated with a laser (electromagnetic wave) of frequency $\omega$ whose magnetic component of field strength $B$ affects only the spin-dynamics of the system. Apart from its connection to the electrodes we also subject the system to onsite dephasing. As we shall show, the magnitude of the onsite dephasing has no effect of the DTC(DTQC) sub-harmonic frequency. The Hamiltonian of the system is described in the methods section. We start by analyzing the Markovian dynamics of the open Fermi-Hubbard chain within the Lindblad master equation \cite{Lindblad, GKS, Breuer, Archak_Dhar_Kulkarni_PhysRevA}, see methods for further details.

 \subsection*{Weak local Floquet-dynamical symmetry} In absence of system-lead coupling (i.e., $\gamma_{L} = \gamma_{R} =0$) the appearance of the TC in the system is stabilized by the presence of a Floquet-dynamical symmetry (FDS) \cite{Neufeld2019,diss_Flq_DTC_PRL_Ikeda, Buca_nat_comm2019, Buca_prb_2020,Buca_PhysRevLettQG_2019, Prosen_SciPostPhys_2020, Tindall_2020}, defined as follows. Given a {Floquet map $\hat{\U}_{F} = \mathcal{T}\left( \text{exp} \left[ \int_{0}^{T} \hat{\mathcal{L}}_{s} ds  \right]\right)$ [and $\hat{\U}_{F}^{-1} = \mathcal{T}\left( \text{exp} \left[- \int_{0}^{T} \hat{\mathcal{L}}_{s} ds  \right]\right)$ where $\mathcal{T}$ is time ordering] if there exists an operator $A$ (in this case it is the total spin raising operator $S^{+} =\sum_{j} S_{j}^{+}$) which satisfies $\hat{A}(T) = \hat{\U}_{F} \hat{A} \hat{\U}_{F}^{-1} = e^{i\lambda T} \hat{A}$, and all the Lindblad operators $V_{\mu}$ satisfy $[V_{\mu}, A(t)] = [V_{\mu}^{\dagger}, A(t)] = 0$, then the system exhibits the FDS and $A(t)$ oscillates in the long-time limit. We follow the notations: $\hat{A}$ represents a super-operator which operates on the vectorized density matrix $|\rho\rangle \rangle$, and $A$ is a normal operator which acts on the desnity matrix as $A \rho$ in the form of a matrix multiplication, where the density matrix $\rho$ satisfies Lindblad equation, \eqref{Lindblad}.} While the oscillations in $A(t)$ were shown to be protected against dephasing \cite{diss_Flq_DTC_PRL_Ikeda}, this is not the case for system-lead coupling; this form of dissipation breaks the FDS. %

Even if the FDS is not fully protected against the system-lead coupling, a DTC behavior can still emerge as a transient phenomenon. To demonstrate this, consider the ``local" operator $S_{loc}^{+} = \left( S_{1}^{+}+S_{N}^{+} \right)$, which satisfies
\begin{equation}\label{weak_dynamical_sym}
    (\U_{F})^{p} (S_{loc}^{+} \rho )  =  e^{i p \lambda T (1 + i \gamma/\lambda)} S_{loc}^{+} (\U_{F})^{p}(\rho),
\end{equation}
 [see Supplementary Note 1 for a proof] after $p$ driving periods for density matrix $\rho$ satisfying Lindblad equation \eqref{Lindblad}, given $[V_{L(R)}, S_{loc}^{+}] \propto \gamma$ ($\gamma_{L} = \gamma_{R} = \gamma$ is assumed). In the long time limit (corresponding to $p\rightarrow \infty$) \eqref{weak_dynamical_sym} represents an approximate FDS as long as $\gamma \ll \lambda$. We call it a ``weak local Floquet dynamical symmetry" (weak local FDS) an emergent FDS in the long time limit near FSS (see Supplementary Note 2).  The notion of the locality comes from the fact that $S_{loc}^{+} = \left( S_{1}^{+}+S_{N}^{+} \right) = ([S_{1}^{+} \otimes I_{2} \otimes \cdot \cdot \cdot \otimes I_{N}] +  [I_{1} \otimes I_{2} \otimes \cdot \cdot \cdot \otimes S_{N}^{+}])$ is local because this is diagonal in the site-basis. The notion of weak dynamical symmetry comes from the fact that $[V_{L(R)}, S_{loc}^{+}] (\propto \gamma)  \neq 0$. The local (unitary) operator $S_{loc}^{+}$ satisfies $\U_{F} ([S_{loc}^{+}]^{m} \rho [S_{loc}^{+ \dagger}]^{n}) = e^{i(m-n) \lambda T} e^{-(m+n)\gamma T} [S_{loc}^{+}]^{m} \U_{F} (\rho )[S_{loc}^{+ \dagger}]^{n}$, (see Supplementary Note 1 for a proof) a Floquet analogue of the weak dynamical condition put forward in Ref. \onlinecite{Prosen_Buca_2012}, see Supplementary Note 3. For $\gamma \approx \lambda$ the FDS is indeed not preserved during the time evolution in the sense that any oscillation in the observable would decay in their respective amplitudes. However, the frequency of the oscillation remains unchanged even at $\gamma \approx \omega$ (see Fig. \ref{fig_phase}).

In the Floquet basis  {(i.e., in the rotating frame)} the equation of motion for $S^{+}_{loc}$ can be shown to be generated by a Floquet Lindbladian $\mathcal{L}_{F}$ whose coherent part is governed by a Floquet Hamiltonian $\mathcal{H}_{F} = \mathcal{H}_{0} + \mathbf{h}\cdot \mathbf{S}$, where $\mathbf{h}\cdot \mathbf{S}$ is a Zeeman term arising from an effective homogeneous and static magnetic field, $\mathbf{h} = (B,0,\omega)$ at each site [see Supplementary Note equations S10]. Given such a Zeeman term we construct, from the local operator $S_{loc, |\mathbf{h}|}^{+} = (S_{1, |\mathbf{h}|}^{+} +S_{N, |\mathbf{h}|}^{+} )$ in the Floquet basis (rotating the axis of quantization along $\mathbf{h}$), a set of coherent states $\rho_{m,n} = (S_{loc, |\mathbf{h}|}^{+})^{m} \rho_{\text{FSS}} (S_{loc, |\mathbf{h}|}^{-})^n$ (with integer values of $m,n$) from the Floquet steady state $\rho_{\text{FSS}}$, satisfying,
\begin{equation}\label{floquet_coherent_states}
    \mathcal{U}_{F} (\rho_{m,n}) = e^{i(m-n) \lambda T} e^{-(m+n)\gamma T} \rho_{m,n},
\end{equation}
 where $\lambda = |\mathbf{h}| = \sqrt{B^2 + \omega^2}$ (modulo $\omega$), and $\rho_{m,n}$ are the Floquet coherent states with an oscillatory component (with a characteristic energy scale $\lambda$) and a decaying part (with a characteristic energy scale $\gamma$). Here $\lambda$ satisfies the usual FDS structure \cite{diss_Flq_DTC_PRL_Ikeda}, viz., if $\lambda/\omega = q/p$ with co-prime integers $p\geq 2$ and $q$, a DTC emerges with a period $T_{DTC}=pT$, otherwise DQTC emerges, and $\gamma/\lambda$ determines the decay of the of the DTC (and DTQC).

For $\gamma \neq 0$, \eqref{floquet_coherent_states} shows that the Floquet steady state (FSS) corresponds to $m=n=0$, i.e., $\rho_{0,0} = \rho_{\text{FSS}}$, and the rest of the states corresponding to $m=n$ are purely decaying states  {(in the sense that these do not exhibit any coherent part)}. However, for $\gamma = 0$ there exists multiple FSS corresponding to $m=n$ \cite{diss_Flq_DTC_PRL_Ikeda}.
 {Clearly for $m\neq n$ and $\gamma = 0$, \eqref{floquet_coherent_states} indicates $\rho_{m,n}$ are degenerate eigenstates of $\U_{F}$ since the same value of $(m-n)$ can be obtained from different combinations of $m,n$-pair. Finite value of $\gamma$ lifts this degeneracy.}

A DTC density matrix (as well as DQTC) can be written as $\rho_{DTC} = \rho_{FSS} + \sum_{\substack{m,n \\ m,n\neq0}} c_{m, n} \rho_{m, n}$, { where $c_{m,n} = |\text{Tr}[\rho_{\text{in}}^{\dagger} \rho_{m,n}]|$, $\rho_{\text{in}}$ being the density matrix of the initial state, are the real} coefficients of the superposition. For $\gamma \ll \lambda$, only a few coherent states $\rho_{m,n}$ decay sufficiently slowly and most of the other coherent states (corresponding to larger integer values of $n$) decay in the long time limit.  {The time evolution of $\rho_{DTC}$ is obtained by $\mathcal{U}(t) \rho_{DTC} [\mathcal{U}(t)]^{-1}$. In the long time limit, the time dependent dissipative DTC density matrix is given by, $\rho_{DTC}(t) = \rho_{FSS} + \sum_{\substack{m,n \\ m,n\neq0}} \left( c_{m, n} e^{i (m-n)\lambda t} \rho_{m, n} + h.c \right)e^{-(m+n)\gamma t}$. Then, observable value of any operator $S^{\alpha} = \prod_{j} S^{\alpha}_{j}$, corresponding to local operator $S^{\alpha}_{j}$ acting on $j'$th site, follows from $\rho_{DTC}(t)$ as $\langle S^{\alpha}(t) \rangle = \text{Tr}[S^{\alpha} \rho_{FSS}] +  \text{Tr}[S^{\alpha} \rho_{DTC}(t)]$ $= \sum_{\substack{m,n \\ m,n\neq0}} 2 \text{Tr}[S^{\alpha}\rho_{m, n}]e^{-(m+n)\gamma t} \cos([m-n]\lambda t) +\text{const.}$. For $\gamma = 0$ [no system-lead coupling], $\langle S^{\alpha}(t) \rangle$ exhibits persistent oscillation, and for non zero $\gamma$ it develops a decaying envelop with decay time $\sim \gamma^{-1}$ leading to meta-stable DTC, see Supplementary Note 1.}

\begin{figure}[h!]
\centering
\includegraphics[keepaspectratio=true,scale=0.38]{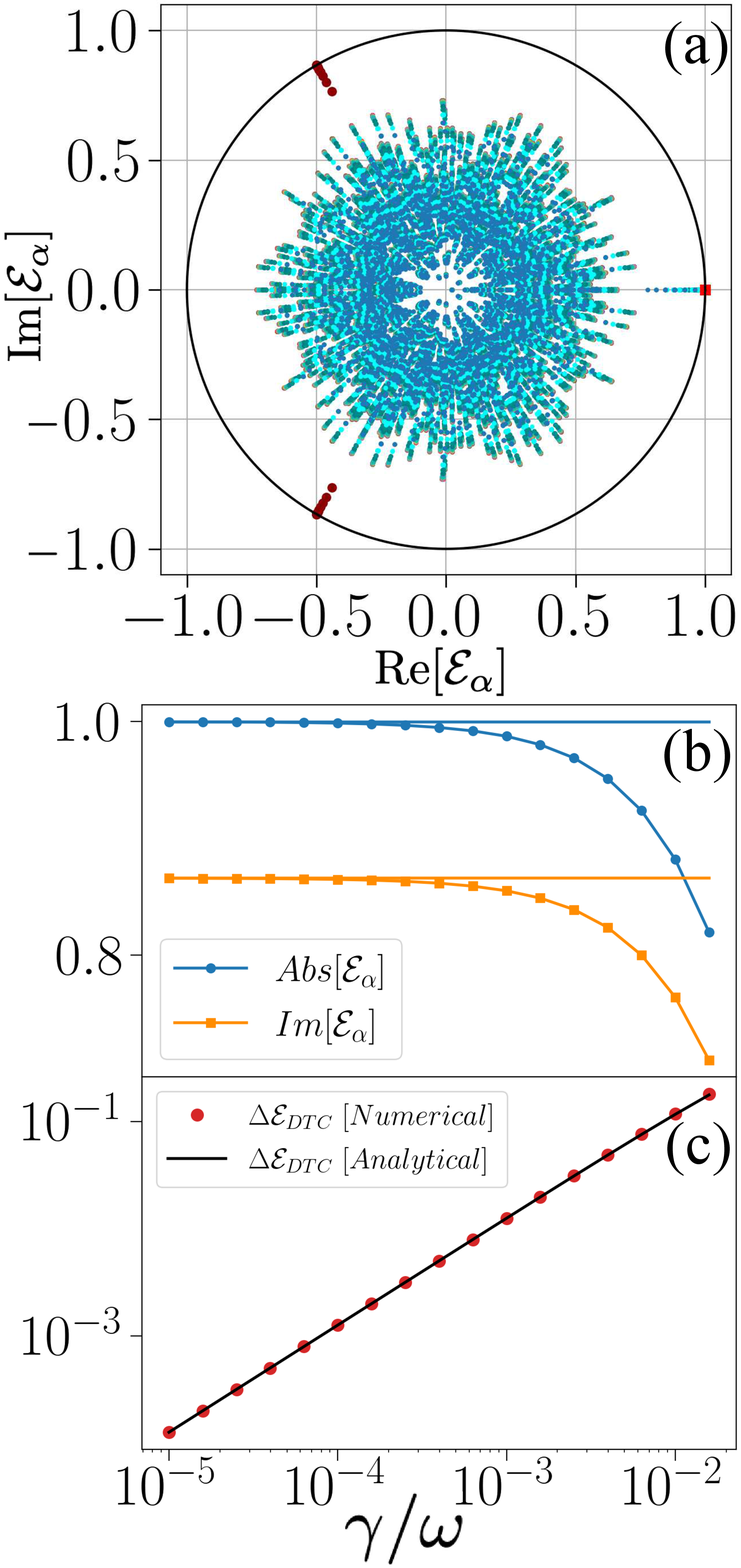}
\caption{{\bf Spectrum of the Floquet map.} (a) All the eigenvalues of the Floquet map $\mathcal{U}_{F}$, the red dots near the unit-circle in the complex plane (not on the real axis) are the discrete time crystal (DTC) eigenvalues. (b) Absolute (Abs) values $\text{Abs}[\E_{DTC}]$ and the imaginary (IM) part $\text{Im}\left[\E_{DTC}\right]$ of the DTC eigenvalues as a function of dimensionless system-lead coupling $\gamma/\omega$. Blue horizontal line represents the peripheral spectrum for which $|\mathcal{E}_{\alpha}| =1$, the orange horizontal line represents $\text{Im}\left[\E_{DTC}\right]$ for $\gamma = 0$. (c) Deviation $\Delta \E_{DTC}$ of the DTC eigen-value $\E_{DTC}$ from the eigenvalue of the Floquet steady state, $\E_{\text{FSS}}$ as a function of $\gamma/\omega$ on a log-log scale. Slope of the straight-line is 0.15. The straight-line corresponds to the function $\Delta \E_{DTC} = \left( 1- e^{-2\gamma T} \right) \approx 2 \gamma T$, where $T=0.1 ~\mu s$ [see Eqs. S(18) - S(22)]. }
\label{fig_Tc_eigs}
\end{figure}

\subsection*{Numerical analysis} 
Following the standard prescription \cite{RieraCampeny2020timecrystallinityin}, we numerically evaluate the eigen-spectrum of the {Floquet map} (taking 3 QDs), viz., $\U_{F} (\rho) = \E_{\alpha} \rho$, $\alpha $ indexing the spectrum and plot $\E_{\alpha}$ on the complex plane in Fig. \ref{fig_Tc_eigs}(a). This eigen-value equation defines the DTC (DQTC) eigen-values as $(\E_{DTC(DTQC)})^{p} = e^{i p \lambda T} = 1$, DTC for integer $p>2$ and DQTC for fractional $p(>2)$, leading the the periodicity of the TC to be $T_{DTC(DQTC)} = p T$, and the FSS eigen-value as $\E_{\text{FSS}} = 1$.  {Fig. \ref{fig_Tc_eigs}(a) further reveals that the meta-stable Floquet coherent states (red dots near the unit circle) are distinct from the FSS (red square on the unit circle) because they never coalesce to the FSS as the system-lead coupling strength increases, indicating the formation of a distinct {transient} manifold \cite{Macieszczak_PhysRevLett_Metastability} within the full spectrum of $\U_{F}$.}

 The system parameters chosen for the above numerical calculations are well within the reach of present day experiments \cite{Barthelemy_2013, Kouwenhoven_2001, Hensgens2017, Mills2019, Zajac439}. We choose $t_{hop}=U=K = 20\pi~ \text{MHz} = 0.26~\mu eV$, all being tunable electostatically, and $\Gamma = 0.1 t_{hop}$. The frequency of the external drive is $\omega = 20 \pi$  {MHz} (therefore, a period of $0.1~\mu s$)  assuming the typical resolution of nano-seconds in the time-dependent measurements in experiments with quantum-dot array, and magnitude of the external magnetic field $B$ is of the order of millitesla.

Corresponding to the specific case of DTC, Fig. \ref{fig_Tc_eigs} (b) shows the absolute value ($\text{Abs}[\E_{DTC}]$) and the imaginary part ($\text{Im}[\E_{DTC}]$) of the DTC eigenvalue $\U_{F}$ as a function of $\gamma$. For sufficiently small values of $\gamma$, $\text{Abs}[\E_{DTC}] = 1$ showing the DTC eigen-value lies on the unit-circle of the spectrum, and $\text{Im}[\E_{DTC}] = \sin \left( \frac{2 \pi l }{3}\right)$ shows that the DTC state has a period $T_{DTC} = 3T$. Stronger system-lead coupling leads DTC eigen-value to move away from the periphery and also introduces a detuning in the DTC time-period as shown in Fig. \ref{fig_Tc_eigs} (b). Therefore, irrespective of the choice of the initial state the spectrum supports the DTC with a period $3T$ as long as $\gamma$ remain small enough, and Fig. \ref{fig_Tc_eigs} (b) corroborates \eqref{weak_dynamical_sym}.

Fig. \ref{fig_Tc_eigs} (c) plots the deviation form the periphery, $\Delta \E_{DTC} = \left( |\E_{\text{FSS}}| - \text{Abs}[\E_{DTC}] \right)$ as function of $\gamma$, showing how the DTC moves away from the peripheral spectrum. Since the least decaying coherent states contribute the most to the DTC, we can therefore identify the DTC density matrix to be  {$\rho_{DTC} =\rho_{FSS} + \frac{1}{2}\left( \rho_{1,0} + \rho_{2,1} + \rho_{0,1} + \rho_{1,2} \right)$}, which is further supported by the numerical calculations shown in Fig. \ref{fig_Tc_eigs} (c) [see supplementary equation S(18) - S(22) for details]. Similarly, {$\rho_{DTC} = \rho_{FSS} + \frac{1}{\sqrt{2}}\left( \rho_{2,0} +\rho_{0,2}\right) $} is also another slowly decaying DTC density matrix that matches the scaling obtained in Fig. \ref{fig_Tc_eigs} (c), [see discussion after supplementary equation S(22)]. However, to which $\rho_{DTC}$ the system will reach depends on the choice of the initial density matrix \cite{Tindall_2020, RieraCampeny2020timecrystallinityin, Gong_PhysRevLett}. 

\begin{figure}[h!]
\centering
\includegraphics[keepaspectratio=true,scale=0.65]{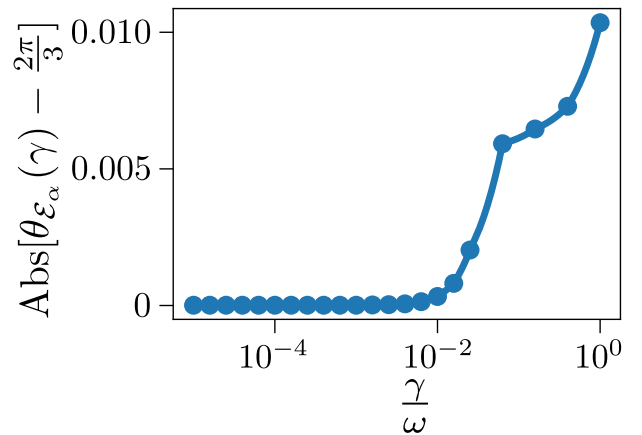}
\caption{{\bf Phase of the discrete time crystal (DTC) eigenvalues relative to $\frac{2\pi}{3}$: } $\text{Abs}[\theta_{\mathcal{E}_{\alpha}}(\gamma) - \frac{2\pi}{3}]$, the absolute (Abs) value of the deviation of the phase $\theta_{\mathcal{E}_{\alpha}}(\gamma) = \arctan \left({\frac{\text{Im}[\mathcal{E}_{\alpha}]}{\text{Re}[\mathcal{E}_{\alpha}]}}\right)$ of DTC eigenvalues $\mathcal{E}_{\alpha}$ from its value $\theta_{\mathcal{E}_{\alpha}}(\gamma=0) =\frac{2\pi}{3}$ at $\gamma = 0$ as a function of $\gamma/\omega$ on a $\log - \log$ scale. }
\label{fig_phase}
\end{figure}

An important consequence of the weak-local FDS is the fact that the DTC oscillation frequency is locked during the time evolution even in presence of system-lead coupling. To highlight this, in Fig. \ref{fig_phase} we plot the phase of the DTC (complex) eigenvalues, $\theta_{\mathcal{E}_{\alpha}}(\gamma) = \arctan \left({\frac{\text{Im}[\mathcal{E}_{\alpha}]}{\text{Re}[\mathcal{E}_{\alpha}]}}\right)$ as a function of $\gamma$. In absence of system-lead coupling $\theta_{\mathcal{E}_{\alpha}}(\gamma = 0) = \frac{2\pi}{3}$ for the specific case of DTC. Fig. \ref{fig_phase} shows that $\theta_{\mathcal{E}_{\alpha}}(\gamma)$ remains locked at $\frac{2\pi}{3}$, and deviates only about $0.5 \%$ even at the largest value of $\gamma = \omega$ we have considered. We have checked that, the deviation is further related to the numerical accuracy corresponding to time of $dt$ of the evolution. In an otherwise perfect numerical calculation ($dt \rightarrow 0$) it would be negligibly small. This frequency locking will further be seen in the oscillation of the spin-current. 

\begin{figure*}
\centering
\includegraphics[keepaspectratio=true,scale=0.25]{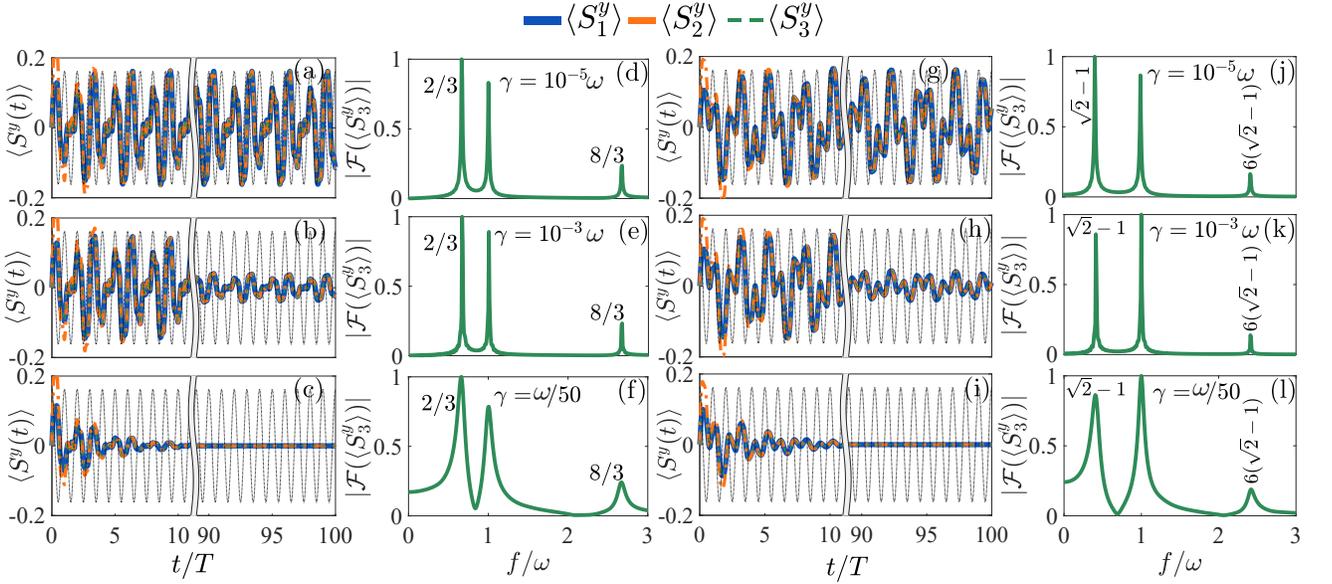}
\caption{{\bf discrete time crystal (DTC) and discrete time quasi-crystal (DTQC) oscillations.} Time evolution of $\langle S^{y}_{j}(t) \rangle$ [$j = 1$ (blue solid lines),$j=2$ (orange dot-dashed lines),$j=3$ (green dashed lines)] as a function of the dimensionless time $t/T$ showing the appearance of DTC in the long-time limit for three values of system-lead couplings, (a) $\gamma = 10^{-5} \omega$, (b) $\gamma = 10^{-3} \omega$, and (c) $\gamma = \omega/50$. Black dotted line represents $\cos (\frac{2\pi t}{T})$, the driving period. (d), (e), (f): Discrete Fourier transform (DTF) $\mathcal{F}(\langle S^{y}_{3} \rangle)$ of $\langle S^{y}_{3}(t) \rangle$ as a function of the frequency domain variable $f/\omega$ corresponding to (a), (b) and (c), respectively, showing the $\lambda = 2\omega/3$. Time evolution of $\langle S^{y}_{j}(t) \rangle$ ($j = 1,2,3$) as a function of the dimensionless time $t/T$ showing the appearance of DTQC in the long-time limit for three values of system-lead couplings, (g) $\gamma = 10^{-5} \omega$, (h) $\gamma = 10^{-3} \omega$, and (i) $\gamma = \omega/50$. (j), (k), (l): DFT of $\langle S^{y}_{3}(t) \rangle$ as a function of the frequency domain variable $f/\omega$ corresponding to (g), (h) and (i), respectively, showing the $\lambda = (\sqrt{2}-1)\omega$. There exists another frequency at an integer multiple of the DTC (and DTQC) frequency in the form of a tiny peak in DTF above $\omega$.
}
\label{fig_spin_osc}
\end{figure*}

\subsection*{Synchronized long-lived DTC and DTQC} 
With the notion of the density matrix corresponding to the DTC (and DTQC) in terms of the decaying coherent states, we numerically demonstrate the role of system-lead coupling in bringing a synchronized stable and meta-stable DTC and DTQC fingerprinted in the oscillation of $\langle S_{j}^{y}(t)\rangle $ ($j=1,2,3$ is the site index) and the spin current. Following the standard notions \cite{Fazio_Quantum_Synchronization, Tindall_2020}, by synchronization we mean that the oscillation of $\langle S_{j}^{y}(t)\rangle $ corresponding to each site is locked to the same frequency and phase, and exhibit the same magnitude independent of the specific value of any microscopic parameters. 

We start from an initial density matrix corresponding to a half filled thermal state at 77K, where the system is placed under a (suitably strong) constant static magnetic field $B_{z}$ in z-direction. Such a configuration is purely chosen as a convenience because it is easy to create in a quantum-dot array. Subsequently, the system is quenched to a state of $B_{z}=0$, and at the same time applied with a circularly polarized magnetic field $V(t)$. However, a random initial density matrix also shows the same results provided it has substantial overlap with the decaying coherent states $\rho_{mn}$. 

Figs. \ref{fig_spin_osc} (a), (b) and (c) show the time evolutions of $\langle S_{j}^{y} \rangle$ in the DTC behavior for $B= \frac{4}{3} \omega$ such that $|\mathbf{h}| = \frac{2}{3}\omega$ (note the modulo operation). Three representative values of $\gamma$ are considered here, corresponding to $\gamma=10^{-5} \omega$ (very weak system-lead coupling) in Fig. \ref{fig_spin_osc} (a), $\gamma=10^{-3} \omega$ (moderate system-lead coupling) in Fig. \ref{fig_spin_osc} (b), and $\gamma=\omega/50$ (strong system-lead coupling) in Fig. \ref{fig_spin_osc} (c), respectively. A system-lead coupling, $10^{-5} \omega \leq \gamma \leq 10^{-3} \omega$ correspond to values in the range of $62~KHz\leq \gamma \leq 6.2~KHz$, which are a standard in transport experiments with quantum-dot-array set-up \cite{Gustavsson_PhysRevLett}.

After an initial relaxation dynamics characterized by the value of the dephasing strength $\Gamma$ ($0.1 t_{hop}$ in our case), $\langle S_{j}^{y} \rangle$ exhibits a stable oscillation with a period $3T$ for weak system-lead coupling, see Fig. \ref{fig_spin_osc} (a). Moreover, the dynamics of $\langle S_{j}^{y} \rangle$ corresponding to all the three sites are synchronized due to the spatial translational symmetry in the density matrix corresponding to the FSS and the other coherent states \cite{Buca_nat_comm2019,  Tindall_2020, diss_Flq_DTC_PRL_Ikeda}. Figs. \ref{fig_spin_osc} (d) shows the corresponding normalized discrete Fourier transform (DFT) exhibiting a peak at $\frac{2}{3}\omega$ commensurately with the driving frequency $\omega$. The secondary peak at $\frac{8}{3}\omega$ is an integer multiple of the DTC peak.

In Fig. \ref{fig_spin_osc} (b), we find that after the initial relaxation $\langle S_{j}^{y} \rangle$ oscillates with decaying magnitude leading to a meta-stable oscillation over a considerable time duration (at-least a few tens of driving periods) before eventually decay to a stationary state with no oscillation. The corresponding DFT is shown in Fig. \ref{fig_spin_osc} (e). This decay occurs in a time scale of $\gamma^{-1}$. This is a manifestation of the fact that for $\gamma = 10^{-3} \omega = 0.063~\text{MHz}$ the DTC eigenvalue of $\mathcal{U}_{F}$ starts to deviate from the peripheral spectrum, see Fig. \ref{fig_Tc_eigs} (c). 

Fig. \ref{fig_spin_osc} (c) further indicates that a stronger damping due to larger values of $\gamma$ leads to a broadened peak in DFT indicating superposition of many frequencies around $\frac{2}{3}\omega$ and $\omega$ signaling a noisy oscillation. This results in broadened DFT peaks seen in Fig. \ref{fig_spin_osc} (f). This is a manifestation of the fact that $\gamma$ acts as a detuning parameter that force $\text{Im} [\mathcal{E}_{DTC}]$ to deviate from its $\gamma=0$ value, see Fig. \ref{fig_Tc_eigs} (c).

A larger system corresponding to N=5 also exhibits the exact same oscillation in $\langle S^{y}_{j}(t) \rangle$ with $3T$ time-period of the DTC, see Supplementary Note 4.

Although dephasing has no effect on the DTC time period, it determines the time scale over which spin-oscillations corresponding to each site get completely synchronized with the other sites. For a given system-lead coupling the first and the last sites are always synchronized irrespective of the value of the dephasing rate. However, all the sites that are not connected to the external leads synchronize on a time scale $\sim \Gamma^{-1}$ with the ones connected to the leads (see Supplementary Note 5). This is true irrespective of the system size, as shown in Supplementary Figure 2 for both four and five sites systems.

The above conclusions hold  true also for the DTQC shown for  $\gamma=10^{-5} \omega$ in Fig. \ref{fig_spin_osc} (g), $\gamma=10^{-3} \omega$ in Fig. \ref{fig_spin_osc} (h), and $\gamma=\omega/50$ in Fig. \ref{fig_spin_osc} (i), respectively. Fig. \ref{fig_spin_osc} (j)-(l) show the corresponding normalized DFTs exhibiting peaks at $f = (\sqrt{2} -1)\omega$ incommensurate with the driving frequency $\omega$. It is worthwhile to point out that if $B $ and $ \omega$ are of the same orders of magnitude or at-least $B>>\omega$, then the DTC/DQTC can be observed. In the case of $B<<\omega$, the coherence frequency scale $\lambda \approx \omega$ and the system follows the conventional Floquet response oscillating with the period of the drive, and no time crystallinity can be observed.
\begin{figure}
\centering
\includegraphics[keepaspectratio=true,scale=0.215]{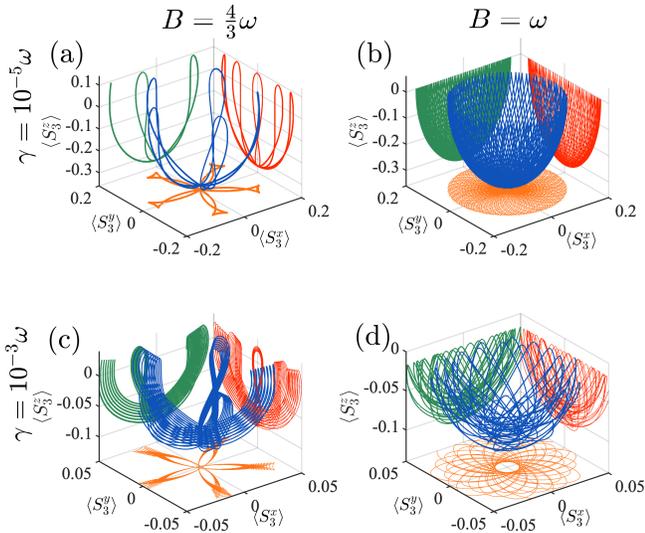}
\caption{{\bf Limit cycles.} Blue lines are trajectories of the time evolution in $\langle S_{3}^{x}\rangle - \langle S_{3}^{y}\rangle -\langle S_{3}^{z}\rangle $ space, (a) and (b) for stable discrete time crystal (DTC) and discrete time quasi-crystal (DTQC), (c) and (d) meta-stable DTC and DTQC, respectively. Orange, green and red lines are projections on $\langle S_{3}^{x}\rangle - \langle S_{3}^{y}\rangle$, $\langle S_{3}^{x}\rangle -\langle S_{3}^{z}\rangle $ and $\langle S_{3}^{y}\rangle -\langle S_{3}^{z}\rangle $ planes, respectively.}
\label{fig_lt_cycle_3d}
\end{figure}

\begin{figure*}
\centering
\includegraphics[keepaspectratio=true,scale=0.2]{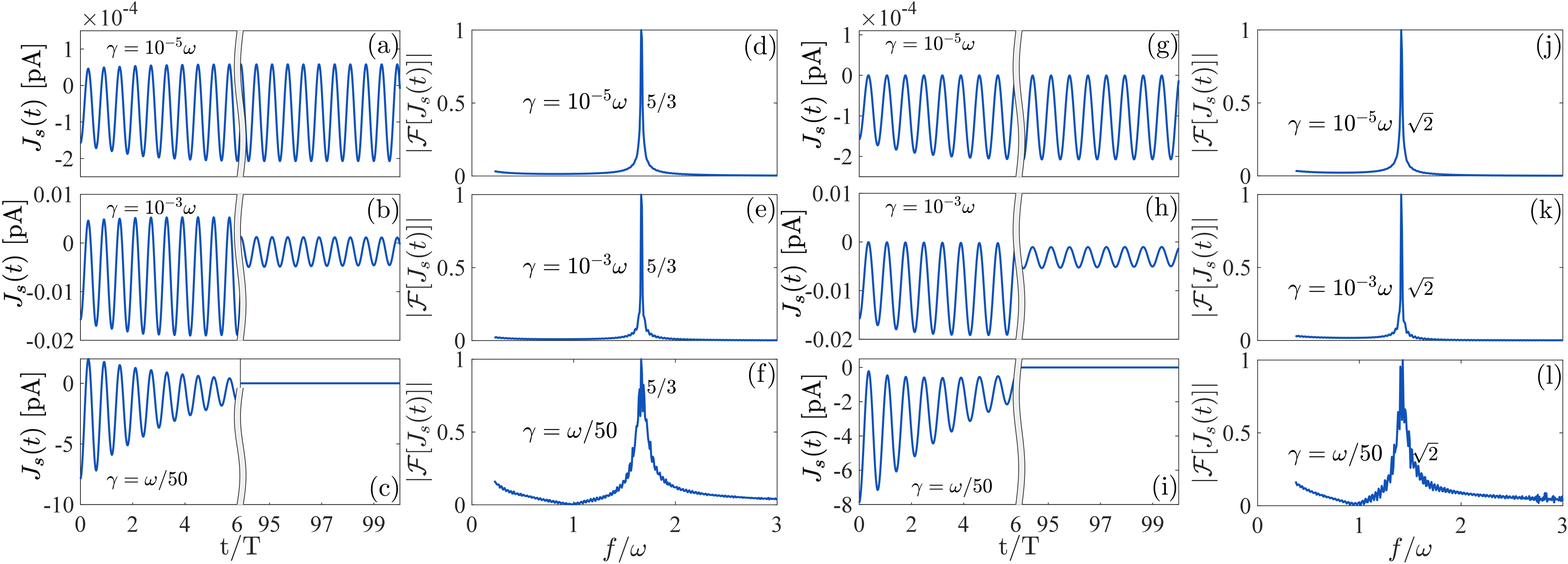}
\caption{{\bf Spin-currents from discrete time crystal (DTC) and time quasi-crystal (DTQC):} Time evolution of the spin currents $J_{s}(t)$, in pico-ampere (pA), as a function of the dimensionless time $t/T$ showing the signature of DTC in the long-time limit for three values of system-lead couplings (a) $\gamma = 10^{-5} \omega$ (very weak system-lead coupling), (b) $\gamma = 10^{-3} \omega$ (moderate system-lead coupling), and (c) $\gamma = \omega/50$ (strong system-lead coupling). (d), (e), (f) correspond to discrete Fourier transform (DFT) $\mathcal{F}[J_{s}(t)]$ of the spin current $J_{s}(t)$ as a function of the frequency domain variable $f/\omega$ corresponding to (a), (b), and (c) respectively, showing the DTC peak at $ f= \frac{5\omega}{3}$. Time evolution of the spin currents $J_{s}(t)$, in pA, as a function of the dimensionless time $t/T$ showing the signature of DTQC in the long-time limit for three values of system-lead couplings (g) $\gamma = 10^{-5} \omega$ (very weak system-lead coupling), (h) $\gamma = 10^{-3} \omega$ (moderate system-lead coupling), and (i) $\gamma = \omega/50$ (strong system-lead coupling). (j), (k), (l) correspond to DFT $\mathcal{F}[J_{s}(t)]$ of the spin current as a function of the frequency domain variable $f/\omega$ corresponding to (g), (h), and (i), respectively, showing the DTQC peak at $ f= \sqrt{2}\omega$.}
\label{fig_spin_current}
\end{figure*}

\subsection*{Limit-cycles}
Although the previous section showed a DTC behavior in $\langle S^{y}(t) \rangle$, we point out that $\langle S^{y}(t) \rangle$ is not easily observed. A connection of the DTC behavior in $\langle S^{y}(t) \rangle$ to a more accessible quantity such as $\langle S^{z}(t) \rangle$, is provided through the limit cycle analysis. A limit cycle is a closed trajectory in the phase-space \cite{pikovsky_rosenblum_kurths_2001}, in our case the phase-space of the dynamical variables, viz., $\langle S_{j}^{x}\rangle$, $\langle S_{j}^{y} \rangle$ and $\langle S_{j}^{z} \rangle$. Fig. \ref{fig_lt_cycle_3d} shows the limit cycle oscillations for both stable and meta-stable DTC and DTQC, respectively.  {To obtain the time evolution we solve \eqref{Lindblad1} [see Methods section] and calculate $\langle S_{j}^{\alpha}(t)\rangle = Tr[S_{j}^{\alpha} \rho(t)]$ for $j=x,y,z$.} The limit cycles are plotted for a time duration of fifty driving periods, for $\gamma=(10^{-5}\omega, 10^{-3}\omega)$ [Fig.\ref{fig_lt_cycle_3d}(a,c)] and $B=(\frac{4}{3}\omega, \omega)$ [Fig.\ref{fig_lt_cycle_3d}(b,d)]. While Figs. \ref{fig_lt_cycle_3d} (a) and (b) show a stable limit cycle characterized by a well defined closed path, \ref{fig_lt_cycle_3d} (c) and (d) show that the limit cycle shrinks ( but very slowly) over fifty driving periods. The limit cycle oscillations indicate that although $S_{j}^{x,(y)}$ respect weak local FDS, $S_{j}^{z}$ also oscillates, providing the means of experimentally extracting a measurable transport signature.

\subsection*{Transport signature-- oscillating spin-current}
Spin current (operator) is defined as,
\begin{equation}
    \hat{J}_{s} = e \gamma (n_{N,\uparrow}-n_{N,\downarrow}) = 2e \gamma S_{N}^{z}, 
\end{equation}
where $e$ is electronic charge, $n_{N,\sigma}$ is the density operator corresponding to the spin $\sigma$ at the extraction site connected to the right lead, see Supplementary Note 6. 
Fig. \ref{fig_spin_current} plots the time evolution of the expectation value of the spin-current, defined above, viz., $J_{s}(t) = \text{Tr} [\hat{J}_{s} \rho(t)]$ and its DFT. In the case of DTC  corresponding to Figs. \ref{fig_spin_current} (a) and (b), a stable and a meta-stable DTC are seen for weak and moderate system-lead couplings, respectively. Importantly enough, the DFT of the corresponding oscillations show only one sharp peak at $f = |\mathbf{h}|$ for both stable and meta-stable DTC, see Figs.  \ref{fig_spin_current} (d) and (e), respectively. This is a direct measurable signature of the DTC in the sense that if the DFT from the spin-current shows a peak at a rational fraction of the driving frequency, i.e.,  $f/\omega = q/p$ one concludes there exists a DTC with a period $pT$. As usual, a strong system-lead coupling destroys the DTC as seen in Fig. \ref{fig_spin_current} (c) and indicated by the broadened peak in the DFT of the spin current in Fig. \ref{fig_spin_current} (f). Figs. \ref{fig_spin_current} (g), (h) and (i) show the oscillation of the spin current corresponding to DTQC for weak, moderate, and strong system-lead couplings, respectively. Likewise, the sharp peak in DFT of the spin current in Figs. \ref{fig_spin_current} (j), (k) and (l) directly provides the exact value of the time period of the DTQC at $f = \sqrt{2} \omega$. The oscillation of spin-current remains exactly the same for the 5-site system, which is apparent from the fact that oscillations in $x-$ and $y-$ components of the spins are exactly the same as that of the 3-site system.

\section*{Conclusions}
We discover, in an experimentally relevant paradigmatic model, viz., Floquet Fermi-Hubbard chain connected to electrodes, DTC and DTQC are manifested through the oscillations of the spin-current. We show that a unique form of ``weak local Floquet dynamical" symmetry emerges in the long time limit which preserves the DTC and the DTQC. The amplitude of the DTC oscillation remain appreciable if the time scale corresponding to the system-lead coupling is much larger than the time period of the drive, viz., $\gamma^{-1} \gg T$. Both DTC and DTQC survive in the long time limit (over 100 periods of external drive) in line with recent experiments performing optical detection of DTCs in closed systems \cite{Kyprianidis1192, mi2021observation}. Due to the ``weak local Floquet dynamical symmetry," a transient/meta-stable manifold emerges which is distinct from the FSS because, with increasing system-electrode coupling, it never coalesces to FSS. DTC and DTQC decay linearly with system-lead coupling strength with slope of the decay being twice the driving period. Our findings thus highlight that by fine-tuning the system-electrode coupling (a form of dissipation engineering) one can obtain a sustained DTC behavior and also allow charge flow through the system allowing measurement of DTC behavior through electrical transport. Put simply, the dynamical symmetry provides the mechanism of DTC/DTQC and meta-stability due to charge transport provides the needed decay channel that aid to measurement
\cite{Volovik2013}. 

Our results show that although the system-lead coupling induces a decay in the oscillation amplitude of DTC, the corresponding sub-harmonic frequency remains locked during the decay, which means that the signature of time-crystallinity can be measured even as it decays. This unfolds an underlying mathematical structure of weak-local Floquet dynamical symmetry. In the long time limit, the system relaxes to a sub-space of of the full super-Hilbert space which is rather non-trivial and can not be a-priori anticipated. Indeed, there can be systems where various effects of environments, such as system-lead coupling, may also lead to the destruction of time-crystal in terms of its oscillation frequency, in which case no signature of the time-crystal can be observed.

Given the pure time-dependent spin current is detectable in the inverse spin-Hall effect \cite{Wei2014}, our predictions are experimentally testable in the presently available quantum-dot array set-up where a Fermi-Hubbard chain has recently been realized \cite{Barthelemy_2013, Kouwenhoven_2001, Hensgens2017, Mills2019, Zajac439,Han2020}. Ultra-cold quantum gases also provide another promising route for experimental study the transport properties of one dimensional many-body systems \cite{Gross995, Brantut1069, Chien2015}. In optical lattices, periodically driven Fermi-Hubbard model is rather routinely analyzed \cite{Esslinger_annurev-conmatphys}, and with the existence of a cold-atom analogue of mesoscopic conductor \cite{Brantut1069, Chien2015} our findings can be verified in optical lattices too. 

\section*{Methods}

\begin{footnotesize}
\subsection*{Hamiltonian}
The Hamiltonian corresponding to periodically driven Fermi-Hubbard chain, see Fig. \ref{fig_set-up}, comprising of $N$ sites is given by $\mathcal{H}(t) = \mathcal{H}_{0} + H_{ext}(t)$ where \cite{diss_Flq_DTC_PRL_Ikeda, Hensgens2017}
\begin{eqnarray}\label{Eq:ham}
\mathcal{H}_{0} &=& -t_{hop} \sum_{\substack{j =1, \\ \sigma = \uparrow,\downarrow}}^{N-1}  (c_{j,\sigma}^{\dagger} c_{j+1,\sigma} + c_{j+1,\sigma}^{\dagger} c_{j,\sigma}) + \sum_{j=1}^{N} \frac{U}{2} n_{j}n_{j} \nonumber \\ &+& \sum_{j=1}^{N-1} \frac{K}{2} n_{j}n_{j+1}; ~ \text{with}~ n_{j} = \sum_{\sigma = \uparrow,\downarrow} n_{j,\sigma}, \nonumber \\
H_{ext}(t) &=& B\sum_{j=1}^{N}(S^{x}_{j} \cos \omega t + S^{y}_{j} \sin \omega t  ).
\end{eqnarray}
 In eq. \eqref{Eq:ham}, $t_{hop},~U,~\text{and}~K$ are the nearest-neighbor hopping, onsite electron-electron, and the nearest-neighbor electron-electron interaction strengths, respectively; $B$ is the magnitude of the external magnetic field induced by a circularly polarized laser of frequency $\omega = \frac{2\pi}{T}$ leading to $\mathcal{H}(t) = \mathcal{H}(t+T)$.

\subsection*{Lindblad equation} 
The system is then subjected to onsite dephasing, represented by the Lindblad operators, $V_{D,j} = \sum_{\sigma} \sqrt{\Gamma} n_{j,\sigma}$ and connected to leads, $V_{L} = \sqrt{\gamma_{L}} \sum_{\sigma=\uparrow, \downarrow}c_{1,\sigma}^{\dagger}$ and $ V_{R} = \sqrt{\gamma_{R}} \sum_{\sigma=\uparrow, \downarrow}c_{N,\sigma}$, where $\gamma_{L(R)}$ is the system- left(right) lead coupling which we take to be equal. The choice of the lead operators mimics the infinite bias condition. The non-unitary dynamics of the system (and the corresponding density matrix $\rho$) is governed by the Floquet-Lindblad equation (within Born-Markov approximation) \cite{Lindblad, GKS, Breuer, Archak_Dhar_Kulkarni_PhysRevA},
 \begin{eqnarray}\label{Lindblad}
 \frac{d \rho}{dt} &=& \mathcal{L}_{t} [\rho] = -i [\mathcal{H}(t), \rho] \nonumber \\ &+& \sum_{\substack{\mu =D ,\\ L,R}} \left(V_{\mu} \rho V_{\mu}^{\dagger} - \frac{1}{2} \lbrace V_{\mu}^{\dagger} V_{\mu}, \rho \rbrace \right),
 \end{eqnarray}
  (we set $\hbar = 1$). The periodicity of the Hamiltonian guarantees the periodicity of the Liouvillian $\mathcal{L}_{t} = \mathcal{L}_{t+T}$.
  
  The density matrix, obeying the Lindblad equation therefore, is an $2^{2N}\times 2^{2N}$ positive definite matrix with trace one. The standard practice of solving the (\ref{Lindblad}) is to vectorize the $\rho$ into a $2^{4N}$ dimensional vector $|\rho \rangle \rangle$. This results in a reformulation of the Lindblad equation into a $2^{4N}$ dimensional Linear ODE with time-dependent coefficients ,
 \begin{equation}\label{Lindblad1}
    \dfrac{|\rho(t)\rangle \rangle}{dt} = \mathcal{L}_{t} |\rho(t) \rangle \rangle,
 \end{equation}
 $\mathcal{L}_{t}$ being the Lindblad super-operator of dimension $2^{4N}\times 2^{4N}$. The infinitesimal time evolution is governed by $|\rho(t+\Delta t)\rangle \rangle = e^{\mathcal{L}_{t} \Delta t} |\rho(t)\rangle \rangle$. 
 
 Alternatively, the Floquet-Lindblad form for any operator $A(t)$ is given by,
\begin{equation}\label{lindblad_eqn_motion}
    \dfrac{d A(t)}{dt} = i[\mathcal{H}(t),A] + \frac{1}{2} \sum_{\mu} \left( V_{\mu}^{\dagger} \left[ A, V_{\mu}\right] + \left[V_{\mu}^{\dagger},  A\right] V_{\mu}\right),
\end{equation}
(notice $\hbar = 1$) where the periodic Hamiltonian $\mathcal{H}(t)$ is given by \eqref{Eq:ham}.
 
 \subsection*{Observables} The oscillations of the spin-components and the spin-current are obtained from the formula, $\langle A(t) \rangle = \text{Tr}[A\rho(t)]$, where $A$ represents the operators whose expectation values are plotted in Figs. \ref{fig_spin_osc} and \ref{fig_spin_current}. The $2^{2N}\times 2^{2N}$ time -dependent density matrix $\rho(t)$ is obtained by reshaping $2^{4N}$ dimensional vector $|\rho(t) \rangle \rangle$.  {Alternatively, we also use Runge-Kutta method for larger system sizes, N=4 and 5. This way of solving the Lindblad equation can be performed in $2^{2N}\times 2^{2N}$ dimensional space.}
 \end{footnotesize}

 \section*{Data availability} All relevant data are available from the corresponding author upon reasonable request.
 
 \section*{Author contribution} SS performed all the analytical and numerical calculations. SS and YD discussed the results and wrote the manuscript. 
 
 \section*{Competing interests} The authors declare no competing interests.

\bibliography{TC_bib}

\end{document}